\documentclass[12pt]{revtex4-1}
\usepackage{amsmath,amssymb}
\usepackage{dcolumn}
\usepackage{bm}
\usepackage[usenames,dvipsnames]{xcolor}
\usepackage{amsthm,amscd}
\usepackage{graphicx}
\usepackage{indentfirst}

\newcommand{\nc}{\newcommand}
\nc{\rnc}{\renewcommand}
\rnc{\[}{\left[}
\rnc{\]}{\right]}
\rnc{\(}{\left(}
\rnc{\)}{\right)}
\nc{\lt}{\left\{}
\nc{\rt}{\right\}}
\rnc{\o}{\omega}
\rnc{\a}{\alpha}
\rnc{\b}{\beta}
\nc{\lam}{\lambda}
\nc{\sig}{\sigma}
\nc{\eps}{\epsilon}
\nc{\vp}{\varphi}
\nc{\del}{\delta}
\nc{\D}{\Delta}
\rnc{\th}{\theta}
\nc{\Th}{\Theta}
\nc{\z}{\zeta}
\nc{\g}{\gamma}
\nc{\ch}{\cosh}
\nc{\sh}{\sinh}
\nc{\bra}{\langle}
\nc{\ket}{\rangle}
\nc{\intall}{\int_{-\infty}^\infty}
\nc{\inth}{\int_0^{\infty}}
\nc{\atan}{\tan^{-1}}
\nc{\sign}{\mbox{sign}}
\nc{\sm}[2]{\sum_{#1}^{#2}}
\nc{\tr}[1]{\mbox{tr}_{#1}}
\nc{\til}{\tilde}
\nc{\w}{\wedge}
\nc{\fr}[2]{\frac{#1}{#2}}
\nc{\obib}[2]{\frac{d {#1}}{d {#2}}}
\nc{\pbib}[2]{\frac{\partial {#1}}{\partial {#2}}}
\nc{\cd}{\cdots}
\nc{\nn}{\nonumber}
\nc{\re}{\eqref}

\begin{document}

\baselineskip 1pt
\begin{flushright}
\begin{tabular}{l}
{\bf OCHA-PP-336}\\
{\bf August 2015}\\
\end{tabular}
\end{flushright}
\baselineskip 24pt
\vglue 10mm

\title{Application of Kawaguchi Lagrangian formulation to string theory}


\author{Ryoko Yahagi}
\email{yahagi@hep.phys.ocha.ac.jp}
\affiliation{Physics Department, Ochanomizu University, 2-1-1 Ohtsuka Bunkyo-ku, Tokyo 112-8610, Japan}

\author{Akio Sugamoto}
\email{sugamoto.akio@ocha.ac.jp}
\affiliation{Physics Department, Ochanomizu University, 2-1-1 Ohtsuka Bunkyo-ku, Tokyo 112-8610, Japan}

\date{\today}


\begin{abstract}

String-scalar duality proposed by Y. Hosotani and membrane-scalar duality by A. Sugamoto are reexamined in the context of Kawaguchi Lagrangian formulation.
The characteristic feature of this formulation is the indifferent nature of fields and parameters.  Therefore even the exchange of roles between fields and parameters is possible.  In this manner, dualities above can be proved easily.  
Between Kawaguchi metrics of the dually related theories, a simple relation is found.
As an example of the exchange between fermionic fields and parameters,
a replacement of the role of Grassmann parameters of the 2-dimensional superspace by the 9th component of Neveu-Schwarz-Ramond (NSR) fermions is studied in superstring model.  Compactification is also discussed in this model.

\end{abstract}

\maketitle

\section{Introduction}

Recently, one of the authors (RY) proposed in collaboration with Ootsuka, Ishida and Tanaka, a covariant Lagrangian formalism
for field theories with the aid of Kawaguchi geometry \cite{oo}.
Kawaguchi space \cite {kawa}, also known as areal space, is defined by a set of a manifold $M$ and an areal metric called Kawaguchi metric $K$ .
In case of field theories, $M$ includes fields $y^i$ $(i=1, \cdots, n)$ and parameters $x^{\mu}$ $(\mu=0, 1, \cdots, D-1)$ of space and time, on the equal footing, and is called extended configuration space.  We label them as $Z^a$ $(a=0, \cdots, n+D-1)$.  Kawaguchi metric $K$ is a kind of Lagrangian density, depending on the coordinates of point in $M$ and its derivatives, 
\begin{eqnarray}
&&K= K (Z^a, ~dZ^{a_1a_2\cdots a_D}),~\mbox{where} \\
&&dZ^{a_1a_2\cdots a_D} \equiv dZ^{a_1} \wedge dZ^{a_2} \wedge \cdots \wedge dZ^{a_D}.
\end{eqnarray}
Integral of $K$ over a given ``sheet" $S$
(or a higher dimensional sub-manifold depending on the theory)
gives an area (or volume) of the ``sheet".  This gives an action of field theories:
\begin{eqnarray}
\mbox{Action}=\int_{S} K
= \mbox{the area or volume of a sub-manifold} ~S.
\end{eqnarray}
To guarantee the reparametrization invariance, the homogeneity condition is imposed on the Kawaguchi metric, namely
\begin{eqnarray}
K (Z, ~\lambda dZ^{a_1a_2\cdots a_D})=\lambda K (Z, ~dZ^{a_1a_2\cdots a_D}).
\end{eqnarray}

In the formulation, fields and parameters are not identified. If we assign $\{y^i\}$ as fields and $\{x^{\mu}\}$ as parameters among $\{Z^a\}$, the fields become functions of the parameters, such as $\sigma:~y^i=y^i(x^0, \cdots, x^{D-1})$.  After the parametrization $\sigma$ is fixed, we have
\begin{align}
\sig^* \( \fr{dx^{0 \cd \mu-1} \w dy^i \w dx^{\mu+1 \cd D-1}}{dx^{0 \cd D-1}} \) = \partial_\mu y^i (x) ,
\end{align}
and the usual description of field theories appears.  This operation $\sigma^{*}$ is sometimes called pullback of a parametrization $\sigma$.  However it is important to note that there are a number of different ways for the parametrization. 

In \cite{oo}, it is proved that every known action can be an area (a volume) of a certain subspace,
so that every field theory can be reformulated \`{a} la Kawaguchi.
Nambu-Goto action is the prototype of this formulation.
The equal treatment of fields and parameters in Kawaguchi Lagrangian formulation gives much potential to reveal dualities which exist between different physical models.

Historically, Finsler introduced in the metric a derivative $\dot{x}^{\mu}$ in addition to the coordinates $x^{\mu}$.  This Finsler metric reads
\begin{eqnarray}
F=F(x^{\mu}, ~\dot{x}^{\mu})dt.
\end{eqnarray}
Physicists understand easily the Finsler metric is nothing but the Lagrangian of quantum mechanics, giving a temporal development of the dynamics in terms of $t$.  Afterwards Kawaguchi generalizes the Finsler geometry so as to include higher order derivatives $\ddot{x}^{\mu}, \cdots$, and also generalize it to the case with many parameters (field theories) mentioned above. 

The purpose of this paper is to derive dualities among different physical models, using the indifferent nature of fields and parameters in the Kawaguchi Lagrangian formalism.

It is known in string theory and membrane theory in four space-time dimensions, the models are dually related to a field theory with two scalars by Hosotani \cite{hoso} and a single scalar field theory by Sugamoto \cite{suga}, respectively.  Therefore we first reexamine these dualities in the next section.  The exchange of fields by parameters is clearly demonstrated. 

A generalization of the dualities given in \cite{hoso} and \cite{suga} is studied by Morris \cite{morris} afterwards.  Baker and Fairlie \cite{b and f} studied scalar field description of $p$-branes, by generalizing the Hamilton-Jacobi formalism of string by Nambu \cite{nambu}.

In section three, we display an example of exchanging fermonic fields and parameters in the superstring model, where a fermionic field (NSR field) in the 9th component $\psi^{9}$ is exchanged by a parameter $\theta$ of the superspace.  Compactification is also discussed in this model.

\section{String and membrane dualities}

In this section, we illustrate how to see dualities in terms of Kawaguchi Lagrangian formalism, taking up two Nambu-Goto type examples.
One example is introduced by Y. Hosotani \cite{hoso}, which gives a duality between Nambu-Goto string
and scalar field theory.
Another example is given by one of the authors (AS) for membrane theory \cite{suga}.
We demonstrate that their dualities can be observed manifestly at action level, which is originally proved by seeing the equations of motion.
Here we consider Euclidean spacetime.

\subsection{String-scalar duality}

The actions for strings and two scalar fields in 4-dimensional spacetime
in \cite{hoso} are given by
\begin{align}
 \label{h1}
 S_{\rm string}
 &=\int d\tau d\sig \sqrt{\fr12 \( V^{\mu\nu}\)^2}, 
 &V^{\mu\nu}
 =\pbib{(X^\mu, X^\nu)}{(\tau, \sig)}, \\
 \label{h2}
 S_{\rm scalars}
 &=\int d^4 x \sqrt{ \fr12 \( W_{\mu\nu} \)^2 },
 &W_{\mu\nu}
 =\pbib{(\rho,\phi)}{(X^\mu, X^\nu)},  
\end{align}
respectively, where $\rho$ and $\phi$ are scalar fields on spacetime $X^\mu, \mu=0,1,2,3$.
$\tau$ and $\sig$ are worldsheet coordinates.
Coefficients are taken arbitrary, since they are of no importance in this argument.
Important fact is that one is 2-dimensional field theory and the other is 4-dimensional field theory.

To consider the duality between these two theories, we set a manifold $M=\{(\tau, \sig, \rho, \phi, X^\mu)\}$.
Kawaguchi metrics for these actions are
\begin{align}
 \label{hk1}
 K_{\rm string}
 &=\sqrt{\fr12 \(dX^{\mu\nu}\)^2 } \\
 \label{hk2}
 K_{\rm scalars}
 &=\sqrt{\fr12 \(dX^{\mu\nu} \w d\rho \w d\phi \)^2 } .
\end{align}
\re{hk1} and \re{hk2} have the same structure; only the difference is the degree of differential forms.
Let $K$ be
\begin{align}
 K(\cdots)=\sqrt{\fr12 (\cdots)^2 }.
\end{align}
We can write
\begin{align}
 S_{\rm string}=\int K \(\pbib{(X^\mu, X^\nu)}{(\xi^0, \xi^1)} \) d\xi^{01},
\end{align}
for arbitrary parametrization $(\xi^0, \xi^1)$.
It can be naturally extended to 4-dimensional field theory by adding extra scalar degrees of freedom as
\begin{align}
 S'=\int K \(\pbib{(X^\mu, X^\nu)}{(\xi^0, \xi^1)} \) d\xi^{01} \w d\rho \w d\phi.
\end{align}
These additional $\rho$ and $\phi$ are degrees of freedom that are perpendicular to the worldsheet.
An identity of Jacobian gives
\begin{align}
 S'&=\int K \(\fr12 \eps^{\mu\nu\lam\eta} \pbib{(X^0, X^1, X^2, X^3)}{(\xi^0, \xi^1, \rho, \phi)}
 \pbib{(\rho, \phi)}{(X^\lam, X^\eta)} \)  d\xi^{01} \w d\rho \w d\phi \nn \\
 &=\int K \(\fr12 \eps^{\mu\nu\lam\eta} \pbib{(\rho, \phi)}{(X^\lam, X^\eta)} \)
 \pbib{(X^0, X^1, X^2, X^3)}{(\xi^0, \xi^1, \rho, \phi)}  d\xi^{01} \w d\rho \w d\phi \nn \\
 &=\int K \( \pbib{(X^\mu, X^\nu, \rho, \phi)}{(X^0, X^1, X^2, X^3)} \) dX^{0123}=S_{\rm scalars},
\end{align}
where $\eps^{\mu\nu\lam\eta}$ is the anti-symmetric Levi-Civita symbol with $\eps^{0123}=1$.
From the first line to the second line of the above equation, we use the homogeneity condition of the Kawaguchi metric.
The last line shows that $S'$ is indeed a pullbacked action determined by \re{hk2}
to the parameter space $(X^0, X^1, X^2, X^3)$.

\subsection{Membrane-scalar duality}

Similar duality can be seen between membrane theory and scalar field theory in 4-dimension.
The actions are
\begin{align}
 \label{s1}
 S_{\rm membrane}
 &=\int d\tau d\sig d\rho \sqrt{\fr{1}{3!} \( V^{\mu\nu\lam}\)^2}, \hspace{10mm} 
 V^{\mu\nu\lam}=\pbib{(X^\mu, X^\nu, X^\lam)}{(\tau, \sig, \rho)}, \\
 \label{s2}
 S_{\rm scalar}
 &=\int d^4 x \sqrt{  \( \pbib{\phi}{X^\mu} \)^2 },
\end{align}
with scalar field $\phi$.

We consider a manifold $M=\{(\tau, \sig, \rho, \phi, X^\mu)\}$, and
Kawaguchi metrics for these actions are
\begin{align}
 \label{sk1}
 K_{\rm membrane}
 &=\sqrt{\fr{1}{3!} \(dX^{\mu\nu\lam}\)^2 } \\
 \label{sk2}
 K_{\rm scalar}
 &=\sqrt{\fr{1}{3!} \(dX^{\mu\nu\lam} \w d\phi \)^2 } .
\end{align}
As well as the string-scalar case, the membrane action \re{s1} is written by 
\begin{align}
 S_{\rm membrane}=\int K \(\pbib{(X^\mu, X^\nu, X^\lam)}{(\xi^0, \xi^1, \xi^2)} \) d\xi^{012}, \hspace{10mm}
 K(\cdots)=\sqrt{\fr{1}{3!} \(\cdots \)^2},
\end{align}
for arbitrary parameters $(\xi^0, \xi^1, \xi^3)$.
Then we obtain
\begin{align}
 S'&=\int K \(\pbib{(X^\mu, X^\nu, X^\lam)}{(\xi^0, \xi^1, \xi^2)} \) d\xi^{012} \w d\phi \nn \\
 &=\int K \(\eps^{\mu\nu\lam\eta} \pbib{(X^0, X^1, X^2, X^3)}{(\xi^0, \xi^1, \xi^2, \phi)}
 \pbib{\phi}{X^\eta} \)  d\xi^{012} \w d\phi \nn \\
 &=\int K \( \pbib{(X^\mu, X^\nu, X^\lam, \phi)}{(X^0, X^1, X^2, X^3)} \) dX^{0123}=S_{\rm scalar}.
\end{align}
It is a pullbacked action of \re{sk2} to the parameter space $(X^0, X^1, X^2, X^3)$.

\vspace{5mm}

At the end of this section, we give a comment on the equivalence of string or membrane model with the scalar model.  Equivalence can be proved by setting the equation of motions in Kawaguchi Lagrangian formulation, and choosing the parametrizations.  In the string-scalar duality, the parametrizaion is $X^{\mu}=X^{\mu}(\tau, \sigma)$ in the string model and $\rho=\rho(X^{\mu})$, $\phi=\phi(X^{\mu})$ in the scalar model, while in the membrane-scalar duality, $X^{\mu}=X^{\mu}(\tau, \sigma, \rho)$ in the membrane model and $\phi=\phi(X^{\mu})$ in the scalar model.  
The result of this section is that the dualities known in \cite{hoso} and \cite{suga} are re-derived manifestly in Kawaguchi Lagrangian formulation, and that a simple relation \re{hk1} and \re{hk2}, or \re{sk1} and \re{sk2} is found between Kawaguchi metrics of dually related theories.  Here the degrees of forms of the Kawaguchi metric is changed by a definite way.

A conjecture at the quantum level on the possible equivalence of string or membrane model with scalar model will be given in the discussion.

\section{Exchange of fermonic fields and variables and compactification}
Kawaguchi Lagrangian formalism has the indifferent nature between fields and variables.  We exemplify the exchange of fermionic fields and variables, taking a superstring model and discuss its compactification.
We start with the superstring action on superspace \cite{Pol},
\begin{align}
 \label{ss}
 S_{\rm superstring}
 &=\fr{1}{4\pi}\int dz d\bar{z} d\th d\bar{\th} \ \bar{{\cal D}} F^\mu {\cal D} F_\mu, \\
 F^\mu
 &=\sqrt{\fr{2}{\a'}} X^\mu (z,\bar{z}) + i \th \psi^\mu (z,\bar{z}) +i \bar{\th} \til{\psi}^\mu (z,\bar{z}), \\
 {\cal D}
 &=\pbib{}{\th}+\th\pbib{}{z},
 \hspace{10mm}
 \bar{{\cal D}}=\pbib{}{\bar{\th}}+\bar{\th}\pbib{}{\bar{z}},
\end{align}
where $\mu=0,1, \cd, 9$.
Complex numbers $z$ and $\bar{z}$ and Grassmann parameters $\theta$ and $\bar{\theta}$ form the 2-dimensional superspace. $X^{\mu}$ gives the location of the string world sheet, and  $\psi^{\mu}$ and $\tilde{\psi}^{\mu}$ gives a two-component spin located on the string world sheet.  The $\alpha^{\prime}$ is a Regge slope parameter and its inverse gives the tension of the string.
Integration with respect to $\th$ and $\bar{\th}$ gives the standard superstring action
\begin{align}
 \label{ss0}
 S_{\rm superstring}
 &=\fr{1}{4\pi}\int dz d\bar{z}
 \( \fr{2}{\a'} \partial X^\mu \bar{\partial} X_\mu
 +\psi^\mu \bar{\partial} \psi_\mu
 +\til{\psi}^\mu \partial \til{\psi}_\mu \),
\end{align}
where $\displaystyle{\partial=\pbib{}{z}}$ and $\displaystyle{\bar{\partial}=\pbib{}{\bar{z}}}$.
Corresponding Kawaguchi space is
\begin{align}
 \label{mani}
 &M={(z, \bar{z}, \th, \bar{\th}, X^\mu, \psi^\mu, \til{\psi}^\mu)} ,\\
 \label{ssk}
 &K_{\rm superstring} 
 =-\fr{(d\xi^{012} \w d F^\mu + d\xi^{023} \w \bar{\th} d F^\mu) (d\xi^{013} \w d F_\mu + d\xi^{123} \w \th d F_\mu)}
 {4\pi \ d\xi^{0123}},
\end{align}
with $(\xi^0, \xi^1, \xi^2, \xi^3)=(z, \bar{z}, \th, \bar{\th})$.
Note that spacetime parameters and fields are on the same footing.
Because of the reparametrization invariance, there is no restriction
that we should pullback the action only to the original spacetime $(z, \bar{z}, \th, \bar{\th})$.
Even $(z, \bar{z}, \psi^9, \til{\psi}^9)$ can be regarded as some other parameter space.
In the latter case fermionic quantities $\th$ and $\bar{\th}$ become functions of $(z, \bar{z}, \psi^9, \til{\psi}^9)$.
For our purpose, we consider the case $\th=\th(\psi^9)=\psi^9, \bar{\th}=\bar{\th}(\til{\psi}^9)=\til{\psi}^9$.
Then the pullback of $F^\mu$ turns into
\begin{align}
 F^\mu=
 \begin{cases}
  \sqrt{\fr{2}{\a'}} X^\mu (z,\bar{z}) + i \th \psi^\mu (z,\bar{z}) +i \bar{\th} \til{\psi}^\mu (z,\bar{z})
  & \mu \neq 9 \\
  \sqrt{\fr{2}{\a'}} X^\mu (z,\bar{z})
  & \mu = 9.
 \end{cases}
\end{align}
Now, the 4-forms appearing in \re{ssk} become
\begin{align}
 d\xi^{0123}
 &=d\z^{0123}  \nn \\
 d\xi^{012} \w d F^\mu
 &=
 \begin{cases}
  d\z^{0123} \( i \til{\psi}^{\mu} \)
  & \mu \neq 9 \\
  0
  & \mu = 9,
 \end{cases} \nn \\
 d\xi^{023} \w \bar{\th} d F^\mu
 &=
 \begin{cases}
  d\z^{0123} \( \z^3 \sqrt{\fr{2}{\a'}} \bar{\partial} X^\mu + \z^3 \z^2 \ i \bar{\partial} \psi^{\mu} \)
  & \mu \neq 9 \\
  d\z^{0123} \( \z^3 \sqrt{\fr{2}{\a'}} \bar{\partial} X^\mu \)
  & \mu = 9,
 \end{cases} \nn \\
 d\xi^{013} \w d F^\mu
 &=
 \begin{cases}
  d\z^{0123} \( -i \psi^{\mu} \)
  & \mu \neq 9 \\
  0
  & \mu = 9,
 \end{cases} \nn \\
 d\xi^{123} \w \th d F^\mu
 &=
 \begin{cases}
  d\z^{0123} \( -\z^2 \sqrt{\fr{2}{\a'}} \partial X^\mu + \z^3 \z^2 \ i \partial \til{\psi}^{\mu} \)
  & \mu \neq 9 \\
  d\z^{0123} \( -\z^2 \sqrt{\fr{2}{\a'}} \partial X^\mu \)
  & \mu = 9,
 \end{cases} \nn \\
\end{align}
where we denote the new parameters as $(\z^0, \z^1, \z^2, \z^3)=(z, \bar{z}, \psi^9, \til{\psi}^9)$.
Finally we obtain
\begin{align}
 \sigma^*_{\zeta} K_{\rm superstring} 
 =&\fr{(d\z^{0123})} {4\pi}
 \biggl[\z^3 \z^2 \( \fr{2}{\a'} \partial X^9 \bar{\partial} X_9
 +\fr{2}{\a'} \partial X^{\mu'} \bar{\partial} X_{\mu'}
 +\psi^{\mu'} \bar{\partial} \psi_{\mu'}
 +\til{\psi}^{\mu'} \partial \til{\psi}_{\mu'} \) \nn \\
 & \hspace{13mm}
 +\z^3 \( i \fr{2}{\a'} \psi^{\mu'} \bar{\partial} X_{\mu'} \)
 +\z^2 \( i \fr{2}{\a'} \til{\psi}^{\mu'} \partial X_{\mu'} \)
 -\til{\psi}^{\mu'} \psi_{\mu'} \biggr],
\end{align}
with $\mu'=0,1, \cd, 8$.
$\sigma^*_{\zeta}$ is inserted to clarify that it is an expression under the parametrization by $\zeta=\{\zeta^{0-3}\}$.
The terms other than the first are indeed 8-dimensional superstrings, 
and the first term shows remaining degree of freedom of compactified space. 

If we integrate the action with respect to $\zeta^2=\psi^9$ and $\zeta^3=\til{\psi}^9$, then we have
\begin{align}\label{sprime}
S_{\rm superstring}^{\prime} 
 =\fr{1}{4\pi}\int dz d\bar{z} \( \fr{2}{\a'} \partial X^9 \bar{\partial} X_9
 +\fr{2}{\a'} \partial X^{\mu'} \bar{\partial} X_{\mu'}
 +\psi^{\mu'} \bar{\partial} \psi_{\mu'}
 +\til{\psi}^{\mu'} \partial \til{\psi}_{\mu'} \) .
\end{align}
The reason why $\theta$ and $\bar{\theta}$ do not recover as fields but disappear from the action after the replacement of the role of $\theta$ and $\bar{\theta}$ with that of $\psi^9$ and $\tilde{\psi}^9$ is in the choice of the parametrization $\sigma_{\zeta}$, where $\theta$ and $\bar{\theta}$ are fixed as $\theta=\psi^9$ and $\bar{\theta}=\tilde{\psi}^9$ without $(z, \bar{z})$ dependence. 
Therefore this choice of parametrization $\sigma_{\zeta}$ triggers the disappearance of $\psi^9$ and $\tilde{\psi}^9$ in \re{sprime}.
The contribution to the action from the 9th direction is 
\begin{align}
\frac{\alpha^{\prime}}{2} \left\{ \left(\frac{m}{R} \right)^2 + \left(\frac{wR}{\alpha^{\prime}} \right)^2 \right\} + \sum_{n=0}^{\infty} n \left(N_n^9+ \tilde{N}_n^9 \right),
\end{align}
if the 9th direction is compactified as a circle with radius $R$.    Correspondingly the first term gives the momentum and winding contributions with Kaluza-Klein excitation number $m$ and winding number $w$, while the second term gives the excitation energy from vibration modes in the 9th direction.  As usual, $N_n^9$ and $\tilde{N}_n^9$ are the occupation number of the n-th vibration modes of right-moving and left-moving modes, respectively.  
The compactification is unfortunately not derived manifestly by the exchange of fermionic fields and variables.

\section{Discussion}

In this paper we have studied two examples, string-scalar duality of \cite{hoso} and membrane-scalar duality of \cite{suga} in the context of Kawaguchi Lagrangian formulation.  In the examples, exchange of fields and parameters is naturally performed owing to the indifferent nature of fields and parameters in Kawaguchi Lagrangian formulation.  Such exchange can be also applied between fermionic fields and fermonic parameters.  Indeed we carry out in the superstring model the exchange between the 9th components ($\psi^9$ and $\tilde{\psi}^9$) of NSR fields and the fermionic coordinates ($\theta$ and $\bar{\theta}$) of the superspace. 

In the proof of dualities in section II, 
a simple relation is found between the Kawaguchi metrics of the dually related theories 
in which the degree of forms in Kawaguchi metric is increased from 2-form to 4-from in the string-scalar duality, from 3-from to 4-from in the membrane-scalar duality, respectively.  This is a very important point.  The equivalence of the models is proved at the classical level, or by showing the equivalence of equations of motion.  Then, what happens if we quantize the models?

Quantization of field theories in Kawaguchi Lagrangian formulation may be given by
\begin{eqnarray}
Z_p= \sum_{S_{p+1}} e^{-\int_{S_{p+1}} K_{p+1}(Z, dZ^{(p+1)}) } ,
\end{eqnarray}
where we explicitly denote the dimensionality $p$ of the configuration which we are studying.  This $p$-dimensionally extended object is now called $p$-brane (If $p=1$ it is string, and if $p=2$, it is membrane.)
In order to quantize $p$-branes, we have to sum over all possible configurations $S_{p+1}$ of $p+1$-dimensional word volume of $p$-branes in $M$.  Let denote the set of $\{S_{p+1}\}$ as ${\cal S}_{p+1}$. 

The N-point amplitude $A_N(x^{(1)}, \cdots, x^{(n)})$ of $p$-branes is discussed in \cite{suga}, in which the amplitude is defined by summing over all the configurations $X^{\mu}$ of $p$-branes which pass through the N-points, $x^{(1)}, \cdots, x^{(N)}$, while the amplitude in the scalar model is defined by summing over all the configurations where each of the scalar fields take the same value at the N-points, $x^{(1)}, \cdots, x^{(N)}$.  This comes from the understanding that the scalar fields ($\rho$ and $\phi$) are the parametrization of space-time, perpendicular to the configuration of string or membrane, so that taking constant values of scalars gives a configuration of string, membrane or $p$-branes in general. 

Therefore, the following conjecture may be given.

Let the extended configuration space $M$ be $M=\{(Z^{a})\}=\{(\xi^{0}, \cdots, \xi^{p}; \xi^{p+1}, \cdots, \xi^{D-1};\allowbreak X^0, \cdots, X^{D-1} )\}$, then a point $P$ has a coordinate $P(\xi^0, \cdots, \xi^{p}; \xi^{p+1}, \cdots,\xi^{D-1}; X^{0}, \cdots, X^{D-1} ) $.  Here we separate the parameters into two categories, $(\xi^0, \cdots, \xi^{p})$ and $(\xi^{p+1}, \cdots,\xi^{D-1})$.  In four dimensional string, the first category is the parametrization of string world sheet $(\tau, \sigma)$, and the second category is the set of scalar fields $(\rho, \phi)$.  Similarly, in four dimensional membrane, the first category is $(\tau, \sigma, \rho)$ and the second category is $(\phi)$.  In section II, we have learned that the string can be described in terms of 2-form action over 2-dimensional surface, or the 4-form action over 4-dimensional surface, while membrane can be described by 3-form action or 4-form action over 3-dimensional or 4- dimensional surface.  Therefore, in order to study dualities we have to prepare more general subspaces, describing the configurations of $q$-branes, where $q$ is not necessarily equal to $p$ but can be $p$, $D-1$, and so forth.

In order to estimate the $N$-point amplitude, a special subspaces ${\cal S}_q (x^{(1)}, \cdots, x^{(N)})$ 
$(\subset  {\cal S}_q)$ should be prepared:
\begin{align}
&{\cal S}_q(x^{(1)}, \cdots, x^{(N)}) \nn \\
&\equiv \{S \in {\cal S}_q \vert 
P^{(i)}(\xi_0, \cdots, \xi_p; \xi_{p+1}=c_{p+1}, \cdots, \xi_{D-1}=c_{D-1}; X^{\mu}=x^{(i)\mu}) \in S, 
i=1-N \}, \hspace{10mm}
\end{align}
where $q$ is not necessarily $p$, as was mentioned above even when we are studying $p$-branes.
If the following amplitudes depending on $q$ is defined by
\begin{eqnarray}
A_N(x^{(1)}, \cdots, x^{(N)})_{q}  
\propto \prod_{j=p+1}^{D-1} \int dc_{j} \sum_{S \in {\cal S}_q(x^{(1)}, \cdots, x^{(N)})} e^{-\int_{S} K_{q+1} (Z, dZ^{(q+1)})},
\end{eqnarray}
then we  may have 
\begin{eqnarray}
A_N(x^{(1)}, \cdots, x^{(N)})_{p} \propto A_N(x^{(1)}, \cdots, x^{(N)})_{D-1}.
\end{eqnarray}
In case of string-scalar duality in section II, we have $(A_N)_1 \propto (A_N)_3$, while in membrane-scalar duality we have $(A_N)_2 \propto (A_N)_3$, where $D=4$. 

At present this is only a conjecture.  However, we hope that the method given in this paper may elucidate various dualities existing in physics and mathematics.

\begin{acknowledgments}

The authors would like to thank Dr. T. Ootsuka for guiding them to the Kawaguchi geometry through valuable discussions and for reading the manuscript.

\end{acknowledgments}



\end{document}